\documentclass[twoside]{article}
\usepackage{fleqn,espcrc2}

\usepackage{graphicx}

\newcommand{\AmS}{{\protect\the\textfont2
  A\kern-.1667em\lower.5ex\hbox{M}\kern-.125emS}}

     {\arraycolsep 0.14em\begin{eqnarray}}{\end{eqnarray}}

\title{Supersymmetric radiative corrections
at large $\tan\beta$ 
\thanks{Talk given at {\it 30 Years of Supersymmetry},
Minneapolis, Minnesota, USA, October 13-27, 2000.}
}

\author{Heather E. Logan\\
	Theoretical Physics Department\\
	Fermi National Accelerator Laboratory, Batavia, IL 60510-0500, USA.}
       
\begin{document}

\begin{abstract}
\vskip-6.5cm
\hfill
FERMILAB-Conf-01/010-T
\vskip0.25cm
\hfill hep-ph/0102029
\vskip5.7cm
In the minimal supersymmetric extension of the Standard Model (MSSM),
fermion masses and Yukawa couplings receive radiative 
corrections at one loop from diagrams involving the supersymmetric particles.  
The corrections to the relation between
down-type fermion masses and Yukawa couplings are enhanced by 
$\tan\beta$, which makes them potentially very significant 
at large $\tan\beta$.
These corrections affect a wide range of processes in the MSSM, 
including neutral and charged Higgs phenomenology, rare $B$ meson decays,
and renormalization of the CKM matrix.
We give a pedagogical review of the sources and phenomenological
effects of these corrections.
\vspace{1pc}
\end{abstract}

\maketitle

\section{INTRODUCTION}
\label{sec:intro}
In the minimal supersymmetric extension of the Standard Model (MSSM)
\cite{HaberKane},
radiative corrections involving supersymmetric (SUSY) particles 
modify the tree level relation between fermion
masses and their Yukawa couplings \cite{HRS}. 
In this paper we review the sources and behavior of these SUSY Yukawa 
corrections and describe their phenomenological effects. 

The behavior of the SUSY Yukawa corrections is most easily derived
in the context of an effective field theory (EFT), 
in which we take the low energy effective theory of the MSSM below the 
SUSY scale
to be a two Higgs doublet model (2HDM) and absorb the effects of SUSY
radiative corrections into the parameters of the EFT.
At tree level, the fermion Yukawa couplings and masses arise from the
Lagrangian,\footnote{For a review of the MSSM Higgs sector, see 
Ref.~\cite{MSSMHiggssector}.}
\begin{equation}
	-\mathcal{L}^{\rm tree} = \epsilon_{ij} \left[
	y_b \bar b_R H_1^i Q_L^j + y_t \bar t_R Q_L^i H_2^j \right]
	+ {\rm h.c.}
\end{equation}
where we use third generation quark notation and the Higgs doublets 
are
\begin{eqnarray}
	H_1 &=& \left(\begin{array}{c}
	(v_1 + \phi_1^{0,r} - i \phi_1^{0,i})/\sqrt{2} \\
	-\phi_1^-
	\end{array} \right), \nonumber \\
	H_2 &=& \left(\begin{array}{c}
	\phi_2^+ \\
	(v_2 + \phi_2^{0,r} + i \phi_2^{0,i})/\sqrt{2}
	\end{array} \right).
\end{eqnarray}
The Higgs vacuum expectation values (vevs) are constrained by
$v_1^2 + v_2^2 = 4 M_W^2/g^2$ and their ratio is parameterized
by $v_2/v_1 \equiv \tan\beta$.
The fermion masses arise from replacing the Higgs fields with their 
vevs:
\begin{eqnarray}
	-\mathcal{L}^{\rm tree} &=& y_b \frac{v_1}{\sqrt{2}} \bar b_R b_L
	+ y_t \frac{v_2}{\sqrt{2}} \bar t_R t_L + {\rm h.c.} \nonumber \\
	&=& m_b \bar b_R b_L + m_t \bar t_R t_L + {\rm h.c.}
	\label{eq:treebmass}
\end{eqnarray}
so that $m_b = \sqrt{2} y_b M_W \cos\beta / g$ and
$m_t = \sqrt{2} y_t M_W \sin\beta / g$.
At tree level, down-type fermions receive their masses from
$H_1$ while up-type fermions receive their masses from $H_2$.  This is
a consequence of the holomorphicity of the superpotential in unbroken
SUSY; if SUSY were unbroken, it would be true to all orders in the EFT.

In a model
with more than one Higgs doublet, tree-level flavor-changing neutral
Higgs couplings can be eliminated if the right-handed fermion 
singlets with each value of hypercharge are allowed to couple to only
one Higgs doublet \cite{GlashowWeinbergPaschos}.  
Under this requirement, called ``natural flavor
conservation'', there are two different configurations possible for
the couplings of the two Higgs doublets $H_1$ and $H_2$ of a 2HDM to
quarks:  in the Type I 2HDM all the quarks
couple to $H_1$ and none to $H_2$, while in the Type II 2HDM the 
down-type right-handed quark singlets couple to $H_1$ while the
up-type right-handed quark singlets couple to $H_2$.   
The most general 2HDM without natural flavor conservation, 
in which the right-handed fermion singlets with
each value of hypercharge couple to both Higgs doublets, is called
the Type III 2HDM.
In unbroken SUSY, the Higgs couplings
take the form of a Type II 2HDM.

\begin{figure}
	\resizebox{7.5cm}{!}
	{\includegraphics*[145,474][402,665]{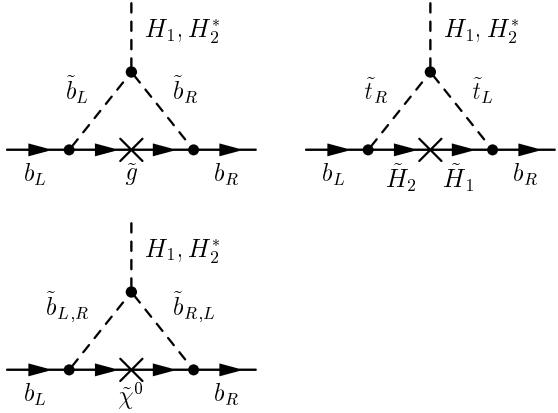}}
	\caption{One-loop diagrams that contribute to Higgs-fermion 
	couplings in broken SUSY.}
	\label{fig:MSSMdiags2}
\end{figure}

However, SUSY must be broken in nature.
When SUSY is broken, non-holomorphic Higgs couplings to fermions
arise at one loop in the EFT.
We first consider down-type fermions.
At one loop, diagrams such as those in Fig.~\ref{fig:MSSMdiags2}
give rise to a correction to the coupling of $b$ quarks to $H_1$ and a
non-holomorphic coupling of $b$ quarks to $H_2$:
\begin{eqnarray}
	-\mathcal{L}^{\rm 1\, loop} &=& 
	\left(y_b + \Delta y_b^{(1)}\right) \bar b_R b_L H_1^0 \nonumber \\
	&& + \left(\Delta y_b^{(2)}\right) \bar b_R b_L H_2^{0*} 
	+ {\rm h.c.}
	\label{eq:effectiveL}
\end{eqnarray}
Examining Eq.~\ref{eq:effectiveL}, we see that in the EFT,
the right-handed down-type quarks
couple to both $H_1$ and $H_2$.
Thus,{\it the low energy effective theory 
of the MSSM is a Type III 2HDM}.\footnote{A review 
of the phenomenology of a 
general Type III 2HDM is given in Ref.~\cite{AtwoodReinaSoni}.}

The $b$ quark mass at one loop is obtained by inserting the Higgs vevs 
into Eq.~\ref{eq:effectiveL}:
\begin{eqnarray}
	-\mathcal{L}^{\rm 1\,loop} &=& 
	\left(y_b + \Delta y_b^{(1)}\right) 
	\frac{v_1}{\sqrt{2}} \bar b_R b_L \nonumber \\
	& & + \left(\Delta y_b^{(2)}\right) 
	\frac{v_2}{\sqrt{2}} \bar b_R b_L + {\rm h.c.}
	\nonumber \\
	&=& m_b \bar b_R b_L + {\rm h.c.}
\end{eqnarray}
Solving for $m_b$ we find,
\begin{eqnarray}
	m_b &=&
	y_b \frac{\sqrt{2} M_W \cos\beta}{g} \nonumber \\
	&& \times \left( 1 + \frac{\Delta y_b^{(1)}}{y_b}
	+ \frac{\Delta y_b^{(2)}}{y_b} \tan\beta \right).
	\label{eq:mbcorr}
\end{eqnarray}
Here $\Delta y_b^{(1,2)}/y_b \ll 1$ 
because of the loop suppression.  
However, the third term in Eq.~\ref{eq:mbcorr} is enhanced
by a factor of $\tan\beta$, and thus can be quite large
at large $\tan\beta$ \cite{HRS,CMWpheno}.
This $\tan\beta$ enhanced correction to $m_b$
has significant phenomenological consequences at large $\tan\beta$
because it modifies the relation between $m_b$ 
and $y_b$.\footnote{In Ref.~\cite{BFPT} it was shown
that it is even possible to have $y_{b,s,d}=0$ at tree level (so that 
$m_{b,s,d}=0$) and generate the down-type quark masses at one loop
through SUSY breaking terms.}
At tree level, $y_b = g m_b/ \sqrt{2} M_W \cos\beta$.
At one loop, 
$y_b$ can be significantly smaller or larger than 
the tree-level value,
depending on the sign of $\Delta y_b^{(2)}/y_b$.

In the up-type fermion sector, an analogous calculation gives,
\begin{eqnarray}
	m_t &=& y_t \frac{\sqrt{2} M_W \sin\beta}{g} \nonumber \\
	&& \times \left( 1 
	+ \frac{\Delta y_t^{(1)}}{y_t} \cot\beta
	+ \frac{\Delta y_t^{(2)}}{y_t} 
	\right).
\end{eqnarray}
In particular, there is no $\tan\beta$ enhancement and the 
corrections to the relation between $m_t$ and $y_t$ are small.
These corrections can
still be important; for example, they enter the MSSM Higgs mass calculation
at the two loop level and lead to a few GeV change in the upper bound
on $M_{h^0}$ in some regions of parameter space \cite{CHHHWW}.

We now examine the behavior of $\Delta y_b^{(1,2)}$
as the SUSY mass scale is varied.
It is easy to see from, e.g., the first diagram in Fig.~\ref{fig:MSSMdiags2}
that $\Delta y_b^{(1,2)}$
are independent of $M_{SUSY}$.  This diagram scales
like $\mu M_{\tilde g} / M^2_{SUSY}$, where the factor of $M_{\tilde g}$
comes from a helicity flip in the gluino line, the factor of $\mu$ comes
from the dimensionful Higgs-squark-squark coupling, and the factor of
$1/M^2_{SUSY}$ comes from the three-point loop integral (here $M_{SUSY}$
is the largest of the masses in the loop and we neglect external momenta;
see Eq.~\ref{eq:Iabc}).
Thus if all the soft SUSY breaking parameters and the $\mu$ parameter
are scaled by a common factor, $\Delta y_b^{(1,2)}$
are unchanged; i.e., they do not decouple with $M_{SUSY}$ when the 
ratios between SUSY parameters are held fixed.
These SUSY Yukawa corrections are also universal; i.e., they are independent
of the external momenta and thus process-independent.
In general, there are additional contributions to physical processes
at one loop that depend on the external momenta; however, these
are suppressed by powers of $p^2/M_{SUSY}^2$.

Although $\Delta y_b^{(1,2)}$ do not decouple with
$M_{SUSY}$, they do not represent a violation of the decoupling
theorem \cite{AppelquistCarazzone}:
in the limit of large pseudoscalar Higgs mass $M_A$, the effects of
$\Delta y_b^{(1,2)}$ appear only in the couplings
of the heavy Higgs bosons, as we will now illustrate.
Because in the Type III 2HDM the right-handed down-type fermions couple
to both Higgs doublets, there is no longer a 
special Yukawa basis for the Higgs fields.  We are then free to write 
them in a new basis chosen such that
one of the doublets has zero vev (the ``vev basis''):
\begin{eqnarray}
	\Phi_1 &=& \left( \begin{array}{c}
	G^+ \\
	(v_{SM} + \Phi_1^{0,r} + iG^0)/\sqrt{2}
	\end{array} \right), \nonumber \\
	\Phi_2 &=& \left( \begin{array}{c}
	H^+ \\
	(\Phi_2^{0,r} + i A^0)/\sqrt{2}
	\end{array} \right),
	\label{eq:vevbasis}
\end{eqnarray}
where $v_{SM} = 2 M_W/g$ is the Standard Model (SM) Higgs vev 
and we have performed the rotation ($H_1^c = i \tau_2 H_1^*$)
\begin{eqnarray}
	H_1^c &=& \cos\beta \Phi_1 - \sin\beta \Phi_2, \nonumber \\
	H_2 &=& \sin\beta \Phi_1 + \cos\beta \Phi_2.
	\label{eq:vevbasisrot}
\end{eqnarray}
Note that in the vev basis $\Phi_1$ contains the Goldstone bosons 
while $\Phi_2$ contains $H^+$ and $A^0$.
The two CP-even neutral Higgs mass eigenstates 
$h^0$ and $H^0$ are given in terms of $\Phi_1^{0,r}$ and $\Phi_2^{0,r}$
by
\begin{eqnarray}
	h^0 &=& \sin(\beta - \alpha) \Phi_1^{0,r} 
	+ \cos(\beta - \alpha) \Phi_2^{0,r} \nonumber \\
	H^0 &=& \cos(\beta - \alpha) \Phi_1^{0,r}
	- \sin(\beta - \alpha) \Phi_2^{0,r}.
\end{eqnarray}
In the limit of large $M_A$, $\cos(\beta - \alpha)$ goes to zero:
\begin{equation}
	\cos(\beta - \alpha) \simeq \frac{M_Z^2 \sin 4\beta}{2 M_A^2},
\end{equation}
so $h^0 \to \Phi_1^{0,r}$.  
Thus in the
limit of large $M_A$, $\Phi_1$ contains the Goldstone bosons, the SM Higgs 
vev, and the light Higgs boson $h^0$, while $\Phi_2$ contains all the heavy
Higgs states, $H^{\pm}$, $A^0$ and $H^0$.  

The effective Lagrangian in the vev basis can be written as,
\begin{equation}
	-\mathcal{L}^{\rm eff} = \lambda_1^b \Phi_1^{0*} \bar b_R b_L
	+ \lambda_2^b \Phi_2^{0*} \bar b_R b_L + {\rm h.c.},
	\label{eq:Leffvevbasis}
\end{equation}
where $\lambda_1^b$ and $\lambda_2^b$ are the Yukawa couplings in the vev
basis.
Inserting the Higgs vevs we obtain the $b$ quark mass:
\begin{eqnarray}
	-\mathcal{L}^{\rm eff} &=& \lambda_1^b \frac{v_{SM}}{\sqrt{2}}
	\bar b_R b_L + {\rm h.c.} \\
	&=& m_b \bar b_R b_L + {\rm h.c.}
\end{eqnarray}
In particular, $\lambda_1^b = \sqrt{2} m_b / v_{SM} = gm_b/\sqrt{2}M_W$ 
is fixed by the $b$ quark mass, while $\lambda_2^b$ is unconstrained.
Thus the coupling of $\Phi_1$ to quarks must be SM-like while that of $\Phi_2$
is unconstrained and can contain the 
effects of $\Delta y_b^{(1,2)}$.
In the decoupling limit \cite{decoupling} of large $M_A$, 
$h^0 \to \Phi_1^{0,r}$ so the $h^0 b \bar b$ coupling approaches its SM value,
while the effects of $\Delta y_b^{(1,2)}$ are
confined to the heavy Higgs bosons in $\Phi_2$.

We now write down the
SUSY radiative corrections to the down-type mass-Yukawa relation
(keeping only the $\tan\beta$ enhanced contributions) \cite{HRS,PBMZ}: 
\begin{eqnarray}
	m_b 
	&\simeq& y_b \frac{\sqrt{2} M_W \cos\beta}{g}
	\left( 1
	+ \frac{\Delta y_b^{(2)}}{y_b} \tan\beta \right) \nonumber \\
	&\equiv& y_b \frac{\sqrt{2} M_W \cos\beta}{g}
	\left( 1 + \Delta_b \right),
	\label{eq:mbyb}
\end{eqnarray}
where the correction $\Delta_b$ is given by~\footnote{Because the SUSY-QCD
contribution to $\Delta_b$ is proportional to the product $\mu M_{\tilde g}$,
it has been suggested in Ref.~\cite{KribsAMSB} to use the sign of $\Delta_b$
and the sign of $\mu$ (measured in some other process) to determine
the sign of $M_{\tilde g}$ and test the anomaly-mediated SUSY breaking
scenario \cite{AMSBthy}, which predicts a negative $M_{\tilde g}$.}
\begin{eqnarray}
	\Delta_b 
	&\simeq& \frac{2 \alpha_s}{3 \pi} M_{\tilde g} \mu \tan\beta \,
	I(M_{\tilde b_1}, M_{\tilde b_2}, M_{\tilde g}) \nonumber \\
	&& + \frac{\alpha_t}{4 \pi} A_t \mu \tan\beta \,
	I(M_{\tilde t_1}, M_{\tilde t_2}, \mu).
	\label{eq:Deltab}
\end{eqnarray}
Here $\alpha_t = y_t^2/4 \pi$ and $I(a,b,c)$ is a loop integral,
\begin{eqnarray}
	I(a,b,c) &=& [a^2 b^2 \log(a^2/b^2) + b^2 c^2 \log(b^2/c^2) 
	\nonumber \\
	& & + c^2 a^2 \log(c^2/a^2)]/ \nonumber \\
	& & [(a^2-b^2)(b^2-c^2)(a^2-c^2)],
	\label{eq:Iabc}
\end{eqnarray}
which is positive for any $a,b,c$ and goes like $1/{\rm max}(a^2,b^2,c^2)$.

Solving Eq.~\ref{eq:mbyb} for $y_b$ \cite{CMWpheno}, we find
\begin{equation}
	y_b = \frac{g m_b}{\sqrt{2} M_W \cos\beta}
	\frac{1}{1 + \Delta_b},
	\label{eq:ybmb}
\end{equation}
where we have not expanded the denominator to one loop order.
It was proven in Ref.~\cite{CGNW} that Eq.~\ref{eq:ybmb}
includes a resummation of terms of order
$(\alpha_s \mu \tan\beta /M_{SUSY})^n$ to all
orders of perturbation theory.
In particular, $\Delta_b$ in Eq.~\ref{eq:ybmb}
gets no $\tan\beta$ enhanced corrections at higher orders 
of the form $(\alpha_s \mu \tan\beta / M_{SUSY})^n$.  The remaining higher
order corrections to $\Delta_b$ are not $\tan\beta$ enhanced and are thus
insignificant compared to the one loop piece (Eq.~\ref{eq:Deltab}).
The proof in Ref.~\cite{CGNW} that Eq.~\ref{eq:ybmb}
receives no higher order $\tan\beta$ enhanced contributions
relies on the facts (1) that $\tan\beta$ enters the calculation at higher
orders only multiplied by $m_b$, e.g., from the $\mu m_b \tan\beta$ term
in the bottom squark mass matrix; and (2) that no factors of $1/m_b$ can
arise from the loop diagrams to cancel the $m_b$ factor multiplying 
$\tan\beta$, because the Yukawa coupling operator is dimension four.

Expanding Eq.~\ref{eq:ybmb} to one loop order, we have
\begin{equation}
	y_b = \frac{g m_b}{\sqrt{2} M_W \cos\beta}
	(1 - \Delta_b).
	\label{eq:ybmb1loop}
\end{equation}
An expression of this form arises in on-shell diagrammatic 
one-loop calculations, in which $\Delta_b$ enters through the 
$b$ quark mass counterterm.  
Although Eqs.~\ref{eq:ybmb} and \ref{eq:ybmb1loop} are equivalent at
one loop, they differ at higher orders:
Eq.~\ref{eq:ybmb1loop} does not contain a resummation of terms of order
$(\alpha_s \mu \tan\beta /M_{SUSY})^n$ to all orders of 
perturbation theory.\footnote{A proof 
as in Ref.~\cite{CGNW} does not hold for Eq.~\ref{eq:ybmb1loop},
in which the $b$ mass operator is renormalized, because the $b$ mass 
operator is not dimension four.  See Ref.~\cite{CGNW} for details.}
In our EFT analyses in the remainder of this paper we will use 
Eq.~\ref{eq:ybmb} in order to take advantage of the resummation of 
higher order terms.

The remainder of this paper is organized as follows.
We examine the effect of the SUSY Yukawa corrections on the process 
$h^0 \to b \bar b$ in Sec.~\ref{sec:hbb} and on
the $H^+ \bar t b$ coupling in Sec.~\ref{sec:H+tb}.  
In Sec.~\ref{sec:FCNCEFT} we
describe how flavor-changing neutral Higgs interactions arise in the EFT
from the SUSY Yukawa corrections, and in Sec.~\ref{sec:FCNCproc} we summarize 
recent results for specific flavor changing processes.  
In Sec.~\ref{sec:CKM} we discuss the renormalization of the CKM matrix.  
In Sec.~\ref{sec:resum} we briefly discuss resummation of the SUSY
Yukawa corrections to flavor changing processes.  
Finally in Sec.~\ref{sec:conclusions} we give a summary and outlook.

\section{SUSY CORRECTIONS TO $\mathbf{h^0 \to b \bar b}$}
\label{sec:hbb}
Accurate knowledge of the $h^0 \to b \bar b$ branching ratio is 
important for determining the reach of the upcoming Higgs searches at the 
Tevatron and LHC \cite{CMWcomplementarity}.  
Also, once a light Higgs boson is discovered, 
precision measurements of its branching ratios (e.g., at a future
$e^+e^-$ linear collider) can be used 
to distinguish a SM Higgs boson from a MSSM Higgs boson 
in some regions of parameter space \cite{CHLM}.
The SUSY Yukawa corrections can have a significant effect on these
branching ratios \cite{CMWpheno,CMWcomplementarity,CHLM,BabuKoldaUnif}.
For example, the ratio $BR(h^0 \to b \bar b)/BR(h^0 \to \tau^+ \tau^-)$ 
is sensitive to the SUSY Yukawa corrections to 
$h^0 \to b \bar b$.\footnote{The $h^0 \tau^+ \tau^-$ coupling also receives
$\tan\beta$ enhanced SUSY Yukawa corrections, but they are proportional
to electroweak gauge couplings and are expected to be much smaller than
the corrections to the $h^0 b \bar b$ coupling.}
At tree level, the ratio $BR(h^0 \to b \bar b)/BR(h^0 \to \tau^+ \tau^-)$
is the same in the SM and the MSSM ($\propto m_b^2/m_{\tau}^2$).  
Further, in the MSSM the $h^0 b \bar b$ and $h^0 \tau^+ \tau^-$ couplings 
have the same dependence on the CP-even Higgs mixing angle $\alpha$, 
so that $BR(h^0 \to b \bar b)/BR(h^0 \to \tau^+ \tau^-)$ is also insensitive
to the radiative corrections to $\alpha$.  
Thus in the context of the MSSM, a deviation of
$BR(h^0 \to b \bar b)/BR(h^0 \to \tau^+ \tau^-)$ from its SM value
provides a direct window onto the SUSY Yukawa corrections
\cite{CMWpheno,CMWcomplementarity,CHLM,BabuKoldaUnif}.

\subsection{EFT calculation}
We begin by computing the SUSY corrections to $h^0 \to b \bar b$ in the 
EFT approach; that is, we assume $M_{SUSY} > M_A$ and neglect 
the external momenta.\footnote{We will consider the effects of the external
momenta and lower $M_{SUSY}$ in the next section.}
Using Eqs.~\ref{eq:effectiveL} and \ref{eq:ybmb}, the 
$h^0 b \bar b$ coupling is given by \cite{CMWpheno}
\begin{eqnarray}
	g_{hbb}^{\rm EFT} &\simeq& \frac{g m_b \sin\alpha}{2 M_W \cos\beta}
	\nonumber \\
	&&\hspace{-0.5cm} \times \left[ 1 - \frac{\Delta_b}{1 + \Delta_b}
	\left(1 + \frac{1}{\tan\alpha \tan\beta} \right) \right]
	\label{eq:ghbbEFT}
\end{eqnarray}
where the tree level coupling is 
$g_{hbb}^{\rm tree} = g m_b \sin\alpha/2 M_W \cos\beta$ and 
$\mathcal{L} = g_{hbb} h^0 \bar b b + \cdots$.
As shown in Sec.~\ref{sec:intro}, the SUSY radiative corrections 
to $g_{hbb}$ should decouple in the limit of large $M_A$.
Indeed, we have,
\begin{equation}
	\left(1 + \frac{1}{\tan\alpha \tan\beta}\right) \simeq
	-\frac{2 M_Z^2}{M_A^2} \cos 2\beta,
\end{equation}
so that $g_{hbb}$ approaches its SM value as expected in the 
decoupling limit \cite{decoupling,HHLPRT}.

\subsection{Diagrammatic calculation}

Up to this point we have neglected the effects of the external 
momenta on $h^0 \to b \bar b$.  These give contributions
of order $(p^2 \tan\beta/M^2_{SUSY})$, where $p^2 = M^2_{h^0}$.  
These contributions can be sizeable because of the
$\tan\beta$ enhancement, especially when the SUSY scale is relatively
low.  For example, for $\tan\beta \sim 50$, 
$(M_{h^0}^2 \tan\beta /M^2_{SUSY})$ is of order one even for 
$M_{SUSY} \sim 1$ TeV.
In order to treat the external momenta correctly, we must do an on-shell
diagrammatic calculation.
We focus here on the SUSY-QCD contributions to $g_{hbb}$,\footnote{The 
SUSY-electroweak contributions will be considered in the next section.}
for which diagrammatic calculations are available in 
Refs.~\cite{HHLPRT,Dabelsteinhbb,Sola}.\footnote{Note that, 
unlike the EFT calculations, 
the one-loop diagrammatic calculations do not incorporate
a resummation of higher order $\tan\beta$ enhanced contributions.}
For compactness we give here an approximate formula\footnote{The full 
one loop result may be found in Refs.~\cite{HHLPRT,Dabelsteinhbb}.} 
expanded in powers
of $M^2_{EW}/M^2_{SUSY}$ and $M^2_{EW}/M^2_A$, where we have
taken $M_{SUSY} \equiv M_S = \mu = M_{\tilde g} = A_b$ and 
$M_S^2 = (M^2_{\tilde b_1} + M^2_{\tilde b_2})/2$.
Defining $g_{hbb}^{\rm diag} = g_{hbb}[1 + \Delta_{SQCD}]$, we have
\cite{HHLPRT}
\begin{eqnarray}
	\Delta_{SQCD}
	&=& \frac{\alpha_s}{3 \pi}
	\left\{ \frac{2 M_Z^2}{M_A^2} \cos 2\beta \tan\beta \right. 
	\nonumber \\
	&& + \frac{M^2_{h^0}}{12 M^2_{SUSY}} (\tan\beta - 1) 
	\nonumber \\
	&& + \frac{M_Z^2}{3 M^2_{SUSY}} \cos 2\beta (\tan\beta - 2) 
	\nonumber \\
	&& + \left. \frac{m_b^2 \tan^2\beta}{2 M^2_{SUSY}} (\tan\beta - 4)
	\right\}.
	\label{eq:dg/g}
\end{eqnarray}
The first term in Eq.~\ref{eq:dg/g} was found before in the EFT calculation;
it remains constant as $M_{SUSY} \to \infty$ at fixed $M_A$.
The remaining terms are neglected in the EFT approach; they decouple
as $M_{SUSY} \to \infty$ but remain 
constant as $M_A \to \infty$ at fixed $M_{SUSY}$.  Because of their 
$1/M^2_{SUSY}$ dependence they can be identified as coming from higher
dimensional operators that were neglected in our EFT calculation.
The correction $\Delta_{SQCD}$ decouples in the limit that both $M_A$
and $M_{SUSY} \to \infty$.


\subsection{SUSY-Electroweak corrections}
The SUSY-electroweak corrections to $h^0 \to b \bar b$
are of order $\alpha_t$ instead of order $\alpha_s$
and arise from the second diagram of Fig.~\ref{fig:MSSMdiags2}.
The SUSY-electroweak Yukawa corrections are easily
calculated in the EFT approach (see Eq.~\ref{eq:Deltab}).
There are also large corrections
to $h^0$ decays at order $\alpha_t$ from the radiative corrections to 
the CP-even Higgs mixing angle $\alpha$ \cite{CMWpheno,Wells}.  
These corrections
are usually taken into account by inserting the
radiatively corrected value of $\alpha$ into the tree level Higgs couplings
to fermions.

At tree level, the off-diagonal elements of the CP-even Higgs mass matrix
are $(\mathcal{M}^2_{12})_{\rm tree} = -(M_A^2 + M_Z^2) \sin\beta \cos\beta$ 
\cite{MSSMHiggssector}.
When $M_A$ is small and $\tan\beta$ is large, $(\mathcal{M}^2_{12})_{\rm tree}$
can be quite small, of the same order as its one loop radiative
corrections.
Then it is possible to tune the radiatively
corrected $\mathcal{M}^2_{12}$ to zero by a careful choice of the MSSM 
parameters, thereby driving the mixing angle $\alpha$ to zero.
If $\sin\alpha = 0$, the
$h^0 b \bar b$ and $h^0 \tau^+ \tau^-$ couplings both vanish at tree level.
However, the SUSY-QCD Yukawa corrections to the
$h^0 b \bar b$ coupling are nonzero, because they depend on the $h^0$
couplings to bottom squarks which remain finite at $\sin\alpha = 0$.
Thus at $\sin\alpha = 0$, the $h^0 \tau^+ \tau^-$
coupling vanishes while the $h^0 b \bar b$ coupling is nonzero due
to the SUSY-QCD Yukawa corrections.  
If we then vary the supersymmetric parameters, we can tune $\alpha$ so that
the tree level value of the $h^0 b \bar b$ coupling cancels its SUSY-QCD 
Yukawa corrections.  Then the corrected $h^0 b \bar b$ coupling vanishes
while the $h^0 \tau^+ \tau^-$ coupling is again nonzero.  Thus the
$h^0 b \bar b$ and $h^0 \tau^+ \tau^-$ couplings vanish at different
values of $\alpha$, or equivalently, at different points in SUSY parameter
space.  
This behavior was first pointed out in Ref.~\cite{CMWcomplementarity}.

Of course,
an on-shell diagrammatic calculation is needed to take into account 
terms of order $(p^2 \tan\beta/M^2_{SUSY})$ from the external momenta.
In Ref.~\cite{HHW} a partial on-shell diagrammatic calculation was
performed, which includes an on-shell calculation
of the corrections to $\alpha$ and the on-shell SUSY-QCD corrections
to $h^0 \to b \bar b$, but does not include the $\mathcal{O}(\alpha_t)$
SUSY-electroweak Yukawa corrections to $h^0 \to b \bar b$.
A full one-loop diagrammatic calculation of the SUSY-electroweak corrections
to $h^0 \to b \bar b$ was performed in Ref.~\cite{Dabelsteinhbb}, although
compact explicit expressions are not available.  A similar calculation
is in progress \cite{HHLPTew}.

\section{SUSY CORRECTIONS TO THE $\mathbf{H^+ \bar t b}$ COUPLING}
\label{sec:H+tb}
The $\tan\beta$ enhanced SUSY Yukawa corrections modify
the $H^+ \bar t b$ coupling through, e.g., 
the first diagram of Fig.~\ref{fig:MSSMdiags2}
with $b_L$ ($\tilde b_L$) replaced by $t_L$ ($\tilde t_L$) and
taking the charged Higgs states in $H_1$, $H_2^*$.
These corrections have important implications for the charged Higgs boson
search through $t \to H^+ b$ at the Tevatron \cite{TevH+tb,GuaschTev,CGNW} 
because
the reach in $M_{H^+}$ and $\tan\beta$
depends on the effective $t$ and $b$ Yukawa couplings.

The one loop SUSY Yukawa corrections to 
$t \to H^+ b$ can be very large, of order the tree level amplitude,
so that reliable calculations require a resummation or improvement.  
Using the EFT approach, we have at large $\tan\beta$ \cite{CGNW}
\begin{equation}
	g_{H^+tb} \simeq \frac{g m_b}{\sqrt{2} M_W} \frac{1}{1 + \Delta_b}
	\tan\beta,
	\label{eq:H+tbresum}
\end{equation}
with $\Delta_b$ as in Eq.~\ref{eq:Deltab}.
As described in Sec.~\ref{sec:intro}, in this result 
$\Delta_b$ gets no $\tan\beta$ enhanced corrections at higher orders 
of the form $(\alpha_s \tan\beta \mu / M_{SUSY})^n$ \cite{CGNW}.  
The remaining higher order corrections to $\Delta_b$ are not 
$\tan\beta$ enhanced.

The SUSY Yukawa corrections to the $H^+ \bar t b$ coupling have
also been computed diagrammatically in
Refs.~\cite{JimenezSola,EHKMY} (SUSY-QCD) and 
\cite{CoarasaSola} (SUSY-electroweak), giving results of the form
\begin{equation}
	g_{H^+tb} = \frac{g m_b}{\sqrt{2} M_W} (1-\Delta_b) \tan\beta.
	\label{eq:H+tbnoresum}
\end{equation}
This is equivalent to Eq.~\ref{eq:H+tbresum} at one loop level,
but in Eq.~\ref{eq:H+tbnoresum} $\Delta_b$ receives 
$\tan\beta$ enhanced radiative corrections at higher orders.
In Ref.~\cite{EHKMY} the tree-level input parameters
were carefully chosen to keep the one-loop corrections in 
Eq.~\ref{eq:H+tbnoresum} small.
This is analogous to using running quark masses to absorb QCD 
corrections; however it does not take into account the higher order
$\tan\beta$ enhanced SUSY corrections.

\section{FLAVOR CHANGING NEUTRAL HIGGS COUPLINGS}
\label{sec:FCNCEFT}
The $\tan\beta$ enhanced SUSY Yukawa corrections lead to flavor changing
neutral Higgs interactions in the EFT.
This was first pointed out in Ref.~\cite{HPT} for
$B$-$\bar B$ mixing and expanded upon in Ref.~\cite{BabuKoldaBmumu}
for leptonic $B$ meson decays.  Similar results have
been found in diagrammatic calculations for a number of
processes; these will be discussed in Sec.~\ref{sec:FCNCproc}.
In this section we describe how flavor changing neutral Higgs 
interactions arise in the EFT.

As shown in Sec.~\ref{sec:intro}, for $M_A < M_{SUSY}$ the low energy
effective theory of the MSSM is the Type III 2HDM.  In the Type III
2HDM, there is no natural flavor conservation and flavor changing 
neutral Higgs couplings arise in general \cite{GlashowWeinbergPaschos}.
In three generation notation, Eq.~\ref{eq:Leffvevbasis} becomes
\begin{equation}
	-\mathcal{L}^{\rm eff} = \lambda_1^{ij} \Phi_1^{0*} \bar d_R^i d_L^j
	+ \lambda_2^{ij} \Phi_2^{0*} \bar d_R^i d_L^j 
	+ {\rm h.c.},
	\label{eq:3genYuks}
\end{equation}
where we again use the vev basis, Eq.~\ref{eq:vevbasis}.
Here $i,j$ are fermion generation indices and $\lambda_{1,2}^{ij}$
are general Yukawa matrices, not necessarily diagonal.
The down-type quark mass matrix is obtained by replacing the Higgs
fields with their vevs:
\begin{equation}
	-\mathcal{L}^{\rm eff} = \lambda_1^{ij} \frac{v_{SM}}{\sqrt{2}} 
	\bar d_R^i d_L^j + {\rm h.c.}
	\label{eq:3genmass}
\end{equation}
Note that $\lambda_2$ does not appear in Eq.~\ref{eq:3genmass} because
$\Phi_2$ has zero vev.
Diagonalizing the mass matrix in Eq.~\ref{eq:3genmass} 
diagonalizes the Yukawa matrix $\lambda_1^{ij}$, which parameterizes
the down-type fermion couplings to $\Phi_1$.
The Yukawa matrix $\lambda_2^{ij}$, which parameterizes
the down-type fermion couplings to $\Phi_2$, is
not in general diagonal in the fermion mass basis.
Rewriting Eq.~\ref{eq:3genYuks} in the fermion mass basis, we have
\begin{equation}
	-\mathcal{L}^{\rm eff} = 
	\frac{g m_i}{\sqrt{2} M_W}
	\Phi_1^{0*} \bar d_R^i d_L^i
	+ \lambda_2^{ij} \Phi_2^{0*} \bar d_R^i d_L^j + {\rm h.c.}
	\label{eq:3gendiag}
\end{equation}
Clearly, the couplings of 
$\Phi_1$ to down-type fermions are flavor diagonal, while those
of $\Phi_2$ are not.  
The off-diagonal elements in $\lambda_2^{ij}$
give flavor changing couplings of the states in $\Phi_2^0$ ($A^0$ 
and $\Phi_2^{0,r}$) to down-type fermions.
Note that the flavor changing
effects of the SUSY Yukawa corrections in physical processes should decouple
at large $M_A$, for which $\Phi_2^{0,r} \simeq H^0$ 
and $M_{H^0} \simeq M_A$.

There are two potential sources of flavor changing neutral Higgs couplings
in the MSSM: minimal and non-minimal flavor violation.
In the MSSM with minimal flavor violation, the sfermion mass matrices
are flavor-diagonal in the same basis as the quark and lepton mass matrices.
Then the only source of flavor changing is the CKM matrix, as in the SM.  
In this case the flavor changing neutral Higgs couplings arise
from one loop diagrams that contain a generation-changing $W^{\pm}$, 
$H^{\pm}$, or chargino coupling and an associated CKM matrix element.
%

Non-minimal flavor violation occurs when the sfermion mass matrices are
not flavor-diagonal in the same basis as the quark and lepton mass
matrices.  
When the sfermion mass matrices are diagonalized,
flavor changing gluino-squark-quark and neutralino-sfermion-fermion
couplings arise. 
Non-minimal flavor violation is present in the
most general MSSM, and can lead to large flavor-changing effects in
contradiction to experiment in, e.g., $K$-$\bar K$ mixing.
This is the SUSY flavor problem, which must be solved in any realistic
MSSM scenario.
Attempts to solve the SUSY flavor problem include
flavor-blind SUSY breaking scenarios (e.g., minimal supergravity),
in which the sfermion mass matrices are flavor diagonal 
in the same basis as the 
quark and lepton mass matrices at the SUSY-breaking scale.  
However,
a small amount of non-minimal flavor violation is typically regenerated 
as the sfermion and fermion mass matrices are run down
to the electroweak scale.
If non-minimal flavor violation is present in the MSSM, 
then the flavor changing neutral Higgs couplings receive contributions
from one loop diagrams that contain a flavor-changing gluino or neutralino
coupling, in addition to the contributions present in the minimal flavor
violation scenario.
%

\section{SOME SPECIFIC FLAVOR CHANGING PROCESSES}
\label{sec:FCNCproc}
\subsection{$b \to s \gamma$}
The SUSY contributions to $b \to s \gamma$ are well known \cite{bsgamma}.
In particular, the chargino-top squark contribution to the 
amplitude is proportional to $\tan\beta$.
It is possible to obey the experimental constraint on $b \to s \gamma$ 
and still have large flavor changing effects in other processes 
from the chargino-top squark loop by choosing heavy enough 
chargino and top squark masses \cite{ChankowskiBmumu}.

Recently in Ref.~\cite{bsgresum} the higher order $\tan\beta$
enhanced contributions to $b \to s \gamma$ were resummed, following
the formalism of Ref.~\cite{CGNW}.
These higher order contributions can be 
large and can modify the regions of MSSM parameter space excluded by
$b \to s \gamma$ \cite{bsgresum}.

\subsection{$B_{s,d} \to \ell^+ \ell^-$}
Because the SM amplitude for $B \to \ell^+ \ell^-$ is helicity suppressed,
contributions from neutral Higgs boson exchange in the 2HDM or MSSM
can compete with the SM amplitude at large $\tan\beta$.
In the MSSM with large $\tan\beta$, 
the dominant contribution to the amplitude comes from 
$h^0$, $H^0$ and $A^0$ exchange with a flavor changing SUSY bubble in the 
external $s$ quark line, and grows with $\tan^3\beta$
\cite{BabuKoldaBmumu,ChankowskiBmumu,Huang00}.
In the MSSM with minimal flavor violation the flavor changing SUSY 
bubble is a chargino-top squark loop; if
non-minimal flavor violation is present,
additional contributions come from a gluino-bottom squark and
neutralino-bottom squark loop.
All other one-loop SUSY contributions have at most a 
$\tan^2\beta$ enhancement.  
In the Type II 2HDM with large $\tan\beta$,
penguin and box diagrams
give a contribution to the amplitude that grows with $\tan^2\beta$
and is of the same order as the SM contribution 
\cite{Huang00,oldBmumu2HDM,Savage,LoganNierste}.

In the MSSM, $B_{s,d} \to \ell^+ \ell^-$ can be enhanced by three
orders of magnitude over the SM rate because of the $\tan^3\beta$
enhancement in the amplitude.
This has been demonstrated
in an EFT calculation in Ref.~\cite{BabuKoldaBmumu} and in diagrammatic
calculations in Refs.~\cite{ChankowskiBmumu,Huang00}.
The experimental sensitivity is best in the channel
$B_s \to \mu^+ \mu^-$; here the SM predicts 
$BR(B_s \to \mu^+ \mu^-) = 4.3^{+0.9}_{-0.8} \times 10^{-9}$ 
\cite{Durhamproc}, while the current experimental bound is
$BR(B_s \to \mu^+ \mu^-) < 2.6 \times 10^{-6}$ (95\% CL) \cite{CDFBmumu}.
At large $\tan\beta$ and low $M_A$, there are regions of SUSY parameter
space in which $BR(B_s \to \mu^+ \mu^-)$ can exceed its current upper bound
\cite{ChankowskiBmumu}; thus SUSY
parameter space is already being probed by this decay.
Future experiments at the Tevatron Run II should improve the 
experimental bound by a factor of 40 with 2 fb$^{-1}$ of data 
\cite{LewisPC}, and further at an extended Run II.
At the LHC this decay is expected to be observed at the SM rate after three
years of running at low luminosity \cite{BallLHC}.

\subsection{$B$-$\bar B$ mixing}

$B$-$\bar B$ mixing has been analyzed in the context of the
effective Type III 2HDM in Refs.~\cite{HPT,BabuKoldaBmumu}.
In the EFT, a tree-level neutral Higgs boson exchange diagram 
with flavor changing couplings at each vertex contributes to 
$B$-$\bar B$ mixing.
If the tree-level relations for MSSM Higgs boson masses and couplings
are used, then the three tree-level EFT diagrams involving $h^0$, $H^0$ 
and $A^0$ exchange sum to zero \cite{BabuKoldaBmumu}.  However, the 
tree-level relations for the MSSM Higgs boson masses and couplings
are in general a poor approximation, and when radiative corrections to
these masses and couplings are included the neutral Higgs mediated 
contributions to $B$-$\bar B$ mixing no longer cancel.

\section{RENORMALIZATION OF THE CKM MATRIX}
\label{sec:CKM}
In the SM, the radiative corrections to the CKM matrix are very small
electroweak effects.  In the MSSM, however,
there are $\tan\beta$ enhanced flavor changing loop diagrams that may
yield significant corrections to the CKM matrix.
In Ref.~\cite{BRP} the down-type quark mass matrix was run down from a 
high scale and the $\tan\beta$ enhanced SUSY corrections were included
at the scale where the SUSY particles were integrated out; diagonalizing
the quark mass matrices then yielded the radiatively corrected CKM 
matrix.
Here we describe the source of the SUSY corrections to the CKM matrix 
in the EFT.

When we diagonalized the down-type quark mass matrix in 
Eq.~\ref{eq:3genmass}, the one loop SUSY corrections were implicitly included.
Writing the one loop corrections explicitly, we have
\begin{eqnarray}
	-\mathcal{L}^{\rm eff} &=& 
	\left(y_{ij} + \Delta y^{(1)}_{ij}\right) 
	H_1^0 \bar d_R^i d_L^j \nonumber \\
	&&+ \left(\Delta y^{(2)}_{ij}\right) 
	H_2^{0*} \bar d_R^i d_L^j + {\rm h.c.}
\end{eqnarray}
In the vev basis this becomes,
\begin{eqnarray}
	-\mathcal{L}^{\rm eff}&&  \nonumber \\
	&&\hspace{-1.5cm}= y_{ij} \cos\beta 
	\left( 1 + \frac{\Delta y^{(1)}_{ij}}{y_{ij}}
	+ \frac{\Delta y^{(2)}_{ij}}{y_{ij}} \tan\beta \right)
	\Phi_1^{0*} \bar d_R^i d_L^j  \nonumber \\
	&&\hspace{-1.2cm} 
	- y_{ij} \sin\beta \left( 1 + \frac{\Delta y^{(1)}_{ij}}{y_{ij}}
	- \frac{\Delta y^{(2)}_{ij}}{y_{ij}} \cot\beta \right)
	\Phi_2^{0*} \bar d_R^i d_L^j  \nonumber \\
	&&\hspace{-1.2cm} + {\rm h.c.}
\end{eqnarray}
Comparing this to Eq.~\ref{eq:3genYuks}, $\lambda_1^{ij}$ is given by
\begin{equation}
	\lambda_1^{ij} = y_{ij} \cos\beta 
	\left( 1 + \frac{\Delta y^{(1)}_{ij}}{y_{ij}}
	+ \frac{\Delta y^{(2)}_{ij}}{y_{ij}} \tan\beta \right).
\end{equation}
Diagonalizing the down-type quark mass matrix diagonalizes $\lambda_1^{ij}$.
Since $\Delta y^{(1,2)}_{ij}$ are not in general diagonal 
in the same basis as the tree-level Yukawa matrix $y_{ij}$, 
they yield an additional rotation of the mass matrix.
This additional rotation is a correction to the CKM matrix.

At present, the CKM elements are not predicted by theory and must
be input from experiment.
If the CKM elements are measured solely through $W$ couplings
to quarks, the SUSY corrections can simply be absorbed into the 
bare elements and are undetectable.
However, if the CKM elements derived from $W$-quark-quark, 
$H^{\pm}$-quark-quark, and especially chargino-squark-quark couplings can
be compared, the SUSY corrections may become apparent.  
We expect the chargino-squark-quark couplings to receive
large corrections relative to the $W$-quark-quark couplings 
because SUSY-breaking effects will lead to different
results for the 
flavor changing SUSY bubble diagrams in the external down-type quark and
down-type squark lines.

\section{RESUMMATION REVISITED}
\label{sec:resum}
The SUSY radiative corrections that lead to flavor changing neutral Higgs 
couplings are enhanced by $\tan\beta$ and can be quite large, as we
have seen.
As in the flavor conserving case, these $\tan\beta$ enhanced corrections 
should be resummed.
Such a resummation has been performed recently in Ref.~\cite{bsgresum}
for higher-order corrections to $b \to s \gamma$, following the
results in Ref.~\cite{CGNW}.
However, it is unclear whether this procedure will work in general because
it is unclear whether the proof in Ref.~\cite{CGNW}
that all higher order $\tan\beta$ enhanced terms are automatically
resummed if the relation in Eq.~\ref{eq:ybmb} is used is applicable to
the flavor changing case.  
For example, factors of $m_b/m_s$ or $m_s/m_b$
can in principle arise at higher orders.  
This issue should be clarified
before the resummation procedure is applied to additional flavor changing
processes.

\section{SUMMARY AND OUTLOOK}
\label{sec:conclusions}
It is clear that the $\tan\beta$ enhanced SUSY corrections to the relation
between down-type fermion masses and their Yukawa couplings has the 
potential to be a good handle on the MSSM at large $\tan\beta$.  
These corrections lead to large effects in both Higgs physics and $B$ physics.
Because these effects can be quite large, reliable calculations are needed.
In Higgs boson decays, the external momenta are
on the order of the electroweak scale and their effects should be included
through diagrammatic calculations; these are available for the SUSY-QCD
corrections to $h^0 \to b \bar b$ and calculations are in progress for
the SUSY-electroweak corrections to both $h^0 \to b \bar b$ and 
$h^0 \to \tau^+ \tau^-$.  
In addition, resummation of higher order 
$\tan\beta$ enhanced contributions beyond the one loop level is very important
in some regions of parameter space.  
This resummation is well known and straightforward
in the EFT approach but should also be incorporated into 
the diagrammatic calculations.  
In flavor changing processes such as
rare $B$ meson decays and CKM matrix renormalization, 
it is not entirely clear how the resummation should be done.  
Finally, the
SUSY corrections to the CKM matrix may have a significant effect
on the $H^{\pm}$ and chargino couplings to (s)fermions 
relative to the $W^{\pm}$ couplings.  
In short, the $\tan\beta$ enhanced SUSY corrections to the relation
between down-type fermion masses and their Yukawa couplings offer a wide
range of challenges and opportunities for MSSM phenomenology.

\vskip0.5cm
{\bf \noindent ACKNOWLEDGMENTS}

I am grateful to the members of the University of Minnesota 
Theoretical Physics Institute for
organizing an excellent symposium and workshop where this work was presented,
and to the other workshop participants for many lively discussions.
I also thank U.~Nierste and T.~Plehn for fruitful discussions.
Fermilab is operated by Universities Research Association Inc.\ 
under contract no.\ DE-AC02-76CH03000 with the U.S.\ Department of
Energy.


\end{document}